\documentclass[12pt,preprint]{aastex}
\newcommand{\myemail}{lixue@xmu.edu.cn}

\begin{document}

\title{Studies of Thermally Unstable Accretion Disks around Black
Holes with Adaptive Pseudo-Spectral Domain Decomposition Method\\
I. Limit-Cycle Behavior in the Case of Moderate Viscosity}

\author{Shuang-Liang Li, Li Xue\altaffilmark{*} and Ju-Fu Lu}
\affil{Department of Physics and Institute of Theoretical Physics
and Astrophysics, Xiamen University, Xiamen, Fujian 361005, China}

\altaffiltext{*}{\myemail}

\begin{abstract}
We present a numerical method for spatially 1.5-dimensional and
time-dependent studies of accretion disks around black holes, that
is originated from a combination of the standard pseudo-spectral
method and the adaptive domain decomposition method existing in
the literature, but with a number of improvements in both the
numerical and physical senses. In particular, we introduce a new
treatment for the connection at the interfaces of decomposed
subdomains, construct an adaptive function for the mapping between
the Chebyshev-Gauss-Lobatto collocation points and the physical
collocation points in each subdomain, and modify the
over-simplified 1-dimensional basic equations of accretion flows
to account for the effects of viscous stresses in both the
azimuthal and radial directions. Our method is verified by
reproducing the best results obtained previously by Szuszkiewicz
\& Miller on the limit-cycle behavior of thermally unstable
accretion disks with moderate viscosity. A new finding is that,
according to our computations, the Bernoulli function of the
matter in such disks is always and everywhere negative, so that
outflows are unlikely to originate from these disks. We are
encouraged to study the more difficult case of thermally unstable
accretion disks with strong viscosity, and wish to report our
results in a subsequent paper.
\end{abstract}


\section{Introduction}
The radiation pressure-supported inner region of geometrically
thin, optically thick Shakura-Sunyaev accretion disks (SSD) around
black holes \citep{SS1973} is known to be thermally unstable
\citep[e.g.][Chap. 4]{Kato98}, but the occurrence of an
instability does not necessarily mean that the disk will be
disrupted after the characteristic growth-time. A possible fate of
the thermally unstable inner region of SSDs is the so-called
limit-cycle behavior, i.e., the nonlinear oscillation between two
stable states. Similar to the case of dwarf novae, the limit-cycle
behavior was realized from a local and steady analysis
\citep[e.g.][Chap.5]{Kato98}, i.e., from an S-shaped sequence of
steady state solutions at a certain radius in the $\dot{M}-\Sigma$
(mass accretion rate vs. surface density) plane, with the lower
and middle branches of the S-shaped sequence corresponding to
stable gas pressure-supported SSD solutions and unstable radiation
pressure-supported SSD solutions, respectively, and the upper
branch corresponding to stable slim disk solutions constructed by
\citet{Abr88}; and has been justified by a number of works
performing global and time-dependent numerical computations
\citep{Honma91,Szuszkiewicz1997,Szuszkiewicz1998,Szuszkiewicz2001,Teresi04a,Teresi04b,Mayer06}.
Unlike the case of dwarf novae, however, only one astrophysical
object, the Galactic microquasar GRS 1915+105, has been known to
show the theoretically predicted limit-cyclic luminosity
variations
\citep{Nayakshin00,Janiuk02,Watarai03,Ohsuga06,Kawata06}.

We select the paper of \citet[][hereafter SM01]{Szuszkiewicz2001}
as the representative of existing theoretical works on the
limit-cycle behavior of black hole accretion disks with the
following two reasons. First, SM01 adopted a diffusion-type
prescription for viscosity, i.e., the $r\phi$ component of the
viscous stress tensor is expressed as
\begin{equation}\label{standar}
\tau_{r\phi}=\alpha H c_s\rho r \frac{\partial \Omega}{\partial
r},
\end{equation}
where $\rho$ is the density, $\Omega$ is the angular velocity,
$c_s$ is the sound speed, $H$ is the half-thickness of the disk,
and $\alpha$ is a dimensionless constant parameter; whereas all
other relevant works used a simple prescription
\begin{equation}\label{alphap}
\tau_{r\phi}=-\alpha p,
\end{equation}
where $p$ is the pressure, and $\alpha$ is also a dimensionless
constant but has been rescaled (denoting $\alpha$ in expressions
[\ref{standar}] and [\ref{alphap}] as $\alpha_1$ and $\alpha_2$,
respectively, then $\alpha_2=[3\sqrt{6}/2]\alpha_1$). It is known
that the direct integration of the differential equations
describing transonic accretion disks with the diffusive form of
viscosity is extremely difficult, while that with the $\alpha p$
viscosity prescription becomes much easier (see the discussion in
SM01). It should be noted, however, that expression (\ref{alphap})
is only an approximation of expression (\ref{standar}) under a
number of conditions \citep[including assuming that the disk is
stationary, geometrically thin, Newtonian Keplerian rotating, and
in vertical hydrostatic equilibrium, e.g.][Chap. 3]{Kato98}. More
seriously, as shown recently by \citet{Becker05}, expression
(\ref{standar}) is the only one proposed so far that is physically
consistent close to the black hole event horizon because of its
diffusive nature, whereas expression (\ref{alphap}) as well as
some other viscosity prescriptions would imply an unphysical
structure in the inner region of black hole accretion disks.
Second, SM01 did complete very nice numerical computations, all
the curves in their figures showing the evolution of disk
structure are perfectly continuous and well-resolved on the grid;
while some fluctuations appear on the curves in the figures of
other relevant works, which might make one to worry whether there
had been some hidden numerical instabilities in the code.

As evidenced by SM01, thermally unstable accretion disks undergo
limit-cycles when viscosity is moderate, i.e., the viscosity
parameter $\alpha\sim0.1$  (hereafter all the numerical values of
$\alpha$ are for $\alpha_2$ unless otherwise specified); and the
instability seems to be catastrophic when viscosity is weak, i.e.,
$\alpha\sim0.001$. On the other hand, in the case of very strong
viscosity, i.e., $\alpha\sim1$, \citet{Chen95} found that the
S-shaped sequence of steady state solutions in the
$\dot{M}-\Sigma$ plane does not form, instead, slim disk solutions
and optically thin advection-dominated accretion flow (ADAF)
solutions \citep{NY94,Abr95} are combined into a single straight
line. Accordingly, \citet{TM98} performed time-evolutionary
computations using the $\alpha p$ viscosity prescription with
$\alpha=1$ and proposed another possible fate of thermally
unstable accretion disks: the very inner region of the disk
finally becomes to be an ADAF-like feature, while the outer region
keeps being the SSD state, forming a persistent two-phased
structure. While this result is really interesting since a
phenomenological SSD+ADAF model has been quite successfully
applied to black hole X-ray binaries and galactic nuclei
\citep[e.g.,][]{Narayan98}, SM01 stated that they could not make
computations for $\alpha=1$ because of difficulties in keeping
their code numerically stable, and pointed out that it is worth
checking whether the persistent ADAF feature obtained in
\citet{TM98} would survive changing the viscosity prescription to
the diffusive form.

We purpose to study thermally unstable accretion disks if they are
not disrupted by instabilities, that is, we wish to check whether
the limit-cycle behavior is the only possible fate of these disks
provided viscosity is not too weak, or a transition from the SSD
state to the ADAF state is the alternative. As in SM01, we adopt
the diffusive viscosity prescription of equation (\ref{standar})
and make spatially 1.5-dimensional, time-dependent computations.
But we choose a numerical method that is different from either of
SM01 or of \citet{TM98}, and that is the adaptive pseudo-spectral
domain decomposition method. With this method, we hope to be able
to perform computations for various values of $\alpha$ ranging
from $\sim0.1$ to $\sim 1$, and to obtain numerical results at the
quality level of SM01. In this paper, we describe our numerical
algorithm and techniques in details and present computational
results for $\alpha=0.1$ as a test of our algorithm. We wish to
report our results for larger values of $\alpha$ in a subsequent
paper.

\section{Numerical Algorithm}
As the main intention of this paper, in this section we present a
numerical algorithm to solve a partial differential equation (or
equations) in the general form
\begin{equation}\label{pde}
    \frac{\partial u(r,t)}{\partial t}=L(u(r,t)),\quad
    r\in[r_{min},r_{max}],
\end{equation}
where $u(r,t)$ is a physical quantity that is a function of the
spatial independent variable $r$ (e.g., the radius in the
cylindrical coordinate system) and the time $t$, and $L$ is a
partial differential operator of $r$ and can be linear or
nonlinear.

\subsection{Scheme of Spacial Discretization}\label{psedo}
We first describe the standard Chebyshev pseudo-spectral method
that is used to discretize the spatial differential operator $L$.
This method has been explained in several textbooks
\citep{gottlieb1983,canuto1988,boyd2000,peyret2002}. Recently,
\citet{Chan2005,Chan2006} applied it to studies of astrophysical
accretion flows and discussed its advantages.

Concretely, a series with finite terms is used to approximate a
physical quantity $u(r)$ as
\begin{equation}\label{appseries}
    u(r_k)=u[g(\bar{r}_k)]=\sum\limits_{n=0}^N\hat{u}_nT_n(\bar{r}_k)=\sum\limits_{n=0}^N\hat{u}_n\cos\left(\frac{nk\pi}{N}\right),
\end{equation}
where $T_n(\bar{r}_k)$ is the $n$-th order Chebyshev polynomial;
$\bar{r}_k$ ($k=0, 1, 2, ... N$) is the Chebyshev-Gauss-Lobatto
collocation points and is defined as $\bar{r}_k\equiv
\cos(k\pi/N)$, with $N$ being the number of collocation points;
$r_k=g(\bar{r}_k)$ is the mapping from the Chebyshev-Gauss-Lobatto
collocation points $\bar{r}_k\in[-1,1]$ to the physical
collocation points $r_k\in[r_{min},r_{max}]$ that is a strictly
increasing function and satisfies both $g(-1)=r_{min}$ and
$g(1)=r_{max}$; $\hat{u}_n$ is the spectral coefficients and can
be calculated from the physical values $u(r_k)$ by a fast discrete
cosine transform \citep[hereafter FDCT,][Chap. 12]{Press1992};
contrarily, if one has $\hat{u}_n$, then $u(r_k)$ can be obtained
immediately by a inverted FDCT.

The radial derivative $\partial u(r)/\partial r$ is also a
function of $r$ and in principle can also be approximated by a
series that is obtained by using the chain rule
\begin{equation}\label{chain}
   \frac{\partial u(r_k)}{\partial
   r}=\frac{1}{dg/d\bar{r}}\frac{\partial u[g(\bar{r}_k)]}{\partial
   \bar{r}}=\frac{1}{dg/d\bar{r}}\sum\limits_{n=0}^N\hat{u}_n^{'}T_n(\bar{r}_k).
\end{equation}
The spectral coefficients $\hat{u}_n^{'}$ can be calculated from
$\hat{u}_n$ by a three-term recursive relation
\begin{eqnarray}\label{three-term}
  \nonumber \hat{u}_N^{'} &=& 0, \\
  \nonumber \hat{u}_{N-1}^{'} &=& 2N\hat{u}_N, \\
  c_n \hat{u}_n^{'} &=& \hat{u}_{n+2}^{'}+2(n+1)\hat{u}_{n+1},
\end{eqnarray}
where $c_0=2$, and $c_n=1$ for $n=1,2,...,N$. Subsequently,
$\partial u[g(\bar{r}_k)]/\partial\bar{r}$ is calculated from
$\hat{u}_n^{'}$ by a inverted FDCT, and then substituted into
equation (\ref{chain}) to obtain discrete spatial derivatives
$\partial u(r_k)/\partial r$.

To summarize, we define a discretized differential operator $D$
for the continuous differential operator $\partial/\partial r$.
The operator $D$ carries out the following works: (1) using FDCT
to calculate $\hat{u}_n$ from $u(r_k)$; (2) using the three-term
recursive relation equation (\ref{three-term}) to obtain
$\hat{u}_n^{'}$ from $\hat{u}_n$; (3) using a inverted FDCT and
equation (\ref{chain}) to obtain $\partial u(r_k)/\partial r$.
Finally, we use $D$ to construct a discretized operator
$\tilde{L}$ to approximate the operator $L$ in equation
(\ref{pde}). For example, if
$L\equiv\partial_r(u_1\partial_ru_2)$, where $\partial_r$ denotes
$\partial/\partial r$, then $\tilde{L}$ can be constructed as
$\tilde{L}=D[u_1D(u_2)]$.

\subsection{Scheme of Time-Discretization}\label{time-discre}
We adopt two schemes to perform the time-integration, that is, we
use a third order total variation diminishing (TVD) Runge-Kutta
scheme \citep{shu1988} to integrate the first two time-steps, and
then change to a low CPU-consumption scheme, the so-called third
order backward-differentiation explicit scheme
\citep[][pp.130-133]{peyret2002}, to carry out the rest
computations.

The third order TVD Runge-Kutta scheme is expressed as
\begin{eqnarray}\label{RK3}
  \nonumber u^{(1)}&=& u^n+\Delta t \tilde{L}(u^n),\\
  \nonumber u^{(2)}&=& \frac{3}{4}u^n+\frac{1}{4}u^{(1)}+\frac{1}{4}\Delta t\tilde{L}(u^{(1)}),\\
  u^{n+1}&=& \frac{1}{3}u^n+\frac{2}{3}u^{(2)}+\frac{2}{3}\Delta
  t\tilde{L}(u^{(2)}),
\end{eqnarray}
where $\Delta t$ is the time-step; $u^n$ and $u^{n+1}$ are the
values of the physical quantity $u$ at the $n$-th and $(n + 1)$-th
time-levels, respectively; and $u^{(1)}$ and $u^{(2)}$ are two
temporary variables.

The third order backward-differentiation explicit scheme can be
written as
\begin{equation}\label{BDE3-1}
    \frac{1}{\Delta t}\sum\limits_{j=0}^{3} a_j
    u^{n+1-j}=\sum\limits_{j=0}^{2} b_j \tilde{L}(u^{n-j}),
\end{equation}
where
\begin{eqnarray}\label{BDE3-2}
  \nonumber a_0&\equiv& 1+\frac{1}{1+k_n}+\frac{1}{1+k_n+k_{n-1}},\\
  \nonumber a_1&\equiv& -\frac{(1+k_n)(1+k_n+k_{n-1})}{k_n(k_n+k_{n-1})},\\
  \nonumber a_2&\equiv& \frac{1+k_n+k_{n-1}}{k_nk_{n-1}(1+k_n)},\\
 a_3&\equiv&
 -\frac{1+k_n}{k_{n-1}(k_n+k_{n-1})(1+k_n+k_{n-1})};\\
 \nonumber\\
  \nonumber b_0&\equiv& \frac{(1+k_n)(1+k_n+k_{n-1})}{k_n(k_n+k_{n-1})},\\
  \nonumber b_1&\equiv& -\frac{1+k_n+k_{n-1}}{k_nk_{n-1}},\\
  b_2&\equiv& \frac{1+k_n}{k_{n-1}(k_n+k_{n-1})};
\end{eqnarray}
and
\begin{eqnarray}\label{BDE3-3}
  \nonumber k_n&\equiv& \frac{t^n-t^{n-1}}{\Delta t},\\
  k_{n-1}&\equiv& \frac{t^{n-1}-t^{n-2}}{\Delta t};
\end{eqnarray}
with $t^n$, $t^{n-1}$, and $t^{n-2}$ being the times of the
$n$-th, $(n-1)$-th, and $(n-2)$-th time-levels, respectively.

Of these two time-integration schemes, the former spends three
times of the latter's CPU-time per time-step, but the latter is
not able to start the time-integration by itself while the former
is able to do. Therefore, we combine these two schemes in order to
achieve a sufficient high order accuracy with minimal CPU-time
consumption.

Hereto, we have fully discretized equation (\ref{pde}). In order
to obtain a physically sound and numerically stable solution in a
finite domain, it is additionally necessary to impose appropriate
boundary conditions and to apply some filtering techniques to
overcome the inevitable spurious nonlinear numerical instabilities
in the code. We leave the details of these in the Appendix.

\subsection{Domain Decomposition}\label{domain-decom}
The numerical algorithm described in the above two subsections and
the Appendix has been a useful implement for solving partial
differential equations and is essentially what was adopted in
\citet{Chan2005}. However, it turns out that, as we have
experienced in our computations, the above algorithm is
insufficient for resolving the so-called stiff problem. This
problem is a one whose solution is characterized by two or more
space-scales and/or time-scales of different orders of magnitude.
In the spatial case, common stiff problems in fluid mechanics are
boundary layer, shear layer, viscous shock, interface, flame
front, etc. In all these problems there exists a region (or exist
regions) of small extent (with respect to the global
characteristic length) in which the solution exhibits a very large
variation \citep[][p. 298]{peyret2002}. When the Chebyshev
pseudo-spectral method described in \S\ref{psedo} is applied to a
stiff problem, the accuracy of the method can be significantly
degraded and there may appear spurious oscillations which can lead
to nonlinear numerical instabilities or spurious predictions of
solution behavior \citep[the so-called Gibbs
phenomenon,][]{gottlieb1997}. The spectral filtering technique
described in the Appendix is not able to completely remove these
spurious oscillations, so that the solution is still not well
resolved, and sometimes the computation can even be destroyed by
the growing spurious oscillations. A special method that has been
developed to overcome these difficulties is the domain
decomposition \citep{bayliss1995,peyret2002}. Here we mainly
follow \citet{bayliss1995} to use this method, but with a
different technique to connect the decomposed subdomains.

The basic idea of domain decomposition is to divide a wide
computational domain into a set of subdomains, so that each
subdomain contains at most only one single region of rapid
variation (i.e., with a stiff problem), and more grid points are
collocated into this region by a special mapping function to
enhance the resolution while the total consumption of CPU-time is
not substantially increased.

In each subdomain the solution is obtained by taking into account
some connection conditions at the interface between the two
conjoint subdomains. In general, appropriate connection conditions
are the continuities of both the solution and its spatial
derivative normal to the interface \citep{bayliss1995,peyret2002}.
The continuity of the solution is satisfied naturally, but the
continuity of its derivative cannot be achieved directly with the
pseudo-spectral method because of the use of FDCT. To see this,
let us divide the entire computational domain
$r\in[r_{min},r_{max}]$ into $M$ subdomains,
\begin{equation}
  S_i\equiv [r_{min}^{(i)},r_{max}^{(i)}],\ \ i=1,2,...,M,
\end{equation}
where $r_{min}^{(1)}=r_{min},\ r_{max}^{(1)}=r_{min}^{(2)},\
r_{max}^{(2)}=r_{min}^{(3)},\ ...$ and $r_{max}^{(M)}=r_{max}$ are
the locations of the interfaces between the subdomains. Because
FDCT is used to calculate the numerical derivative in each
subdomain $S_i$, one obtains two values of the derivative at each
interface. Let $\partial^{-}u$ and $\partial^{+}u$ denote the left
and right numerical derivatives of the physical quantity $u$ at a
certain interface, respectively, a seemingly rational choice for
keeping the continuity of derivative is to set the numerical
derivative at the interface to be the mean of $\partial^{-}u$ and
$\partial^{+}u$, i.e.,
\begin{equation}\label{connect}
   \left(\frac{\partial u}{\partial r}\right)_{interface}=\frac{\partial^{-}u+\partial^{+}u}{2}.
\end{equation}
Unfortunately, in practice the connection technique of equation
(\ref{connect}) will often cause a numerical instability at the
interfaces.

We find that the connection between two certain subdomains $S_i$
and $S_{i+1}$ can be numerically stable when their discretizations
satisfy an additional practical condition. Let
$r_{int}$($=r_{max}^{(i)}=r_{min}^{(i+1)}$) denotes the location
of interface between $S_i$ and $S_{i+1}$; $r_{N-1}^{(i)}$ and
$r_1^{(i+1)}$ ($r_{N-1}^{(i)}\in S_i$, $r_1^{(i+1)}\in S_{i+1}$
and $r_{N-1}^{(i)}<r_{int}<r_1^{(i+1)}$) denote the locations of
the two nearest points to the interface, respectively; our
computations show that if the condition
\begin{equation}\label{connect2}
   \left|r_{int}-r_{N-1}^{(i)}\right|=\left|r_{int}-r_1^{(i+1)}\right|
\end{equation}
is satisfied, then the connection of derivative represented by
equation (\ref{connect}) will be numerically stable.

If the stiff problem always appeared in a fixed spatial region,
then the domain decomposition would be kept unchanged. However, in
general this is not the case. Instead, the location of region in
which the stiff problem appears changes with time
\citep{bayliss1995}. Therefore, the domain decomposition must be
adjusted adaptively. To ensure the connection condition equation
(\ref{connect2}) at the interfaces of newly divided subdomains, an
adjustable mapping between the physical collocation points $r_k$
in each new subdomain $S_i$ and the Chebyshev-Gauss-Lobatto
collocation points $\bar{r}_k$ is needed. We adopt such a mapping
in the form (see eq.[\ref{appseries}])
\begin{equation}\label{adjmapp}
   r^{(i)}=g(\bar{r})=r_{max}^{(i)}+\frac{2}{\pi}\left(r_{max}^{(i)}-r_{min}^{(i)}\right)\arctan\left[a
   \tan\frac{\pi}{4}(\bar{r}-1)\right],\ \ \bar{r}\in [-1,1]\ \mathrm{and}\ r^{(i)}\in
   S_i,
\end{equation}
in the subdomain $S_i$, which is a combination of two mapping
functions,
\begin{equation}\label{trivialmap}
   r^{(i)}=\frac{r_{max}^{(i)}}{2}(\tilde{r}+1)-\frac{r_{min}^{(i)}}{2}(\tilde{r}-1)
\end{equation}
and
\begin{equation}\label{adjmapp2}
   \tilde{r}=\frac{4}{\pi}\arctan\left[a
   \tan\frac{\pi}{4}(\bar{r}-1)\right].
\end{equation}
Equation (\ref{trivialmap}) is a trivial linear mapping
\citep{Chan2005}, and equation (\ref{adjmapp2}) is the mapping
proposed by \citet{bayliss1995}. The parameter $a$ in equation
(\ref{adjmapp2}) is an adjustable parameter, but equation
(\ref{adjmapp2}) is only a mapping from $\bar{r}\in[-1,1]$ to
$\tilde{r}\in[-1,1]$. Therefore, we add equation
(\ref{trivialmap}) in order to make a complete mapping from
$\bar{r}\in[-1,1]$ to $r^{(i)}\in S_i$. The combined mapping
equation (\ref{adjmapp}) will concentrate the discrete grid points
toward $r_{min}^{(i)}$ when $a>1$ and toward $r_{max}^{(i)}$ when
$a<1$, and will be reduced to equation (\ref{trivialmap}) when
$a=1$.

The adjustability of mapping equation (\ref{adjmapp}) is crucially
important for achieving a numerically stable connection at the
interfaces of subdomains. By substituting equation (\ref{adjmapp})
into equation (\ref{connect2}), we obtain
\begin{equation}\label{para-a}
   a^{(i+1)}=\frac{\cot\left\{\omega\arctan\left[a^{(i)}\tan\frac{\pi}{4}(\bar{r}_l-1)\right]\right\}}{\tan\frac{\pi}{4}(\bar{r}_r-1)},
\end{equation}
with
\begin{eqnarray}
  \nonumber
  \omega&\equiv&\frac{r_{max}^{(i)}-r_{min}^{(i)}}{r_{max}^{(i+1)}-r_{min}^{(i+1)}},\\
  \nonumber \bar{r}_l&\equiv&\cos\left(\frac{\pi}{N}\right),\\
  \nonumber \bar{r}_r&\equiv&\cos\left[\frac{(N-1)\pi}{N}\right],
\end{eqnarray}
where $a^{(i)}$ and $a^{(i+1)}$ are the mapping parameters for
subdomains $S_i$ and $S_{i+1}$, respectively. Equation
(\ref{para-a}) can be used to determine the mapping parameters of
every subdomain after giving a decomposition of computational
domain $\{S_i\}$ and the mapping parameter $a^{(1)}$ of the
innermost subdomain $S_1$ ($=[r_{min},r_{max}^{(1)}]$). As a
result, we obtain a particular collocation of discrete grid-points
within the whole computational domain $[r_{min},r_{max}]$. This
collocation ensures a stable connection of the derivatives between
any two conjoint subdomains (eq.[\ref{connect}]), and thus ensures
a correct implementation of the pseudo-spectral method in each
subdomain. The combination of the standard pseudo-spectral method
and the adaptive domain decomposition method finally names our
numerical algorithm as that in the title of this paper.

\section{Limit-cycle Solutions}
We now verify our numerical algorithm by applying it to studies of
thermally unstable black hole accretion disks with moderate
viscosity and comparing our results with that of the
representative work SM01.

\subsection{Basic Equations}
We write basic equations for viscous accretion flows around black
holes in the Eulerian form rather than Lagrangian form as in SM01
because partial differential equations in the Eulerian description
take the general form of equation (\ref{pde}), to which our
numerical algorithm is suited. The basic equations to be solved
are
\begin{eqnarray}
  \frac{\partial \Sigma}{\partial t}&=&-v_r\frac{\partial \Sigma}{\partial
   r}-\frac{\Sigma}{r}\frac{\partial}{\partial
   r}(rv_r),\label{eq-mass}\\
  \frac{\partial v_r}{\partial t}&=&-v_r\frac{\partial v_r}{\partial r}-\frac{1}{\rho}
    \frac{\partial p}{\partial r}+\frac{l^2-l_K^2}{r^3},\label{eq-radial}\\
  \frac{\partial l}{\partial t}&=&-v_r\frac{\partial l}{\partial r}+\frac{\alpha}{r\Sigma}
    \frac{\partial}{\partial r}\left(r^3c_sH\Sigma\frac{\partial\Omega}{\partial r}\right),\label{eq-angular}\\
  \frac{\partial H}{\partial t}&=&-v_r\frac{\partial H}{\partial r}+V_z,\label{eq-height}\\
  \frac{\partial V_z}{\partial t}&=&-v_r\frac{\partial V_z}{\partial r}+6\frac{p}{\Sigma}-\frac{GM_{BH}}{(r-r_g)^2}\left(\frac{H}{r}\right),\label{eq-vertical}\\
  \nonumber\frac{\partial T}{\partial t}&=&-v_r\frac{\partial T}{\partial
  r}\\
   & &+\frac{T}{12-10.5\beta}\left\{\frac{\alpha\Sigma
c_sH(r\partial\Omega/\partial
  r)^2-F^{-}}{0.67pH}-(4-3\beta)\left[\frac{V_z}{H}+\frac{1}{r}\frac{\partial}{\partial
  r}(rv_r)\right]\right\}.\label{eq-energy}
\end{eqnarray}

Equations (\ref{eq-mass}), (\ref{eq-radial}), (\ref{eq-angular}),
and (\ref{eq-energy}) are conservations of mass, radial momentum,
angular momentum, and energy, respectively. As in SM01, we adopt
the diffusive form of viscosity (eq. [\ref{standar}]) in equations
(\ref{eq-angular}) and (\ref{eq-energy}); and abandon vertical
hydrostatic equilibrium assumed in 1-dimensional studies, and
instead introduce two new dynamical equations for the vertical
acceleration (eq. [\ref{eq-vertical}]) and the evolution of the
disk's thickness (eq. [\ref{eq-height}]), thus our studies can be
regarded as 1.5-dimensional. In these equations $v_r$, $l$, $l_K$,
$V_z$, $M_{BH}$, $r_g$, $T$, $\beta$, and $F^{-}$ are the radial
velocity, specific angular momentum, Keplerian specific angular
momentum, vertical velocity at the surface of the disk, black hole
mass, gravitational radius ($\equiv 2GM_{BH}/c^2$), temperature,
ratio of gas pressure to total pressure, and radiative flux per
unit area away from the disk in the vertical direction,
respectively. We use the 'one-zone' approximation of the
vertically-averaged disk as in SM01, so that $v_r$, $\Omega$, $l$,
$l_K$, $\rho$, $p$, $c_s$, and $T$ are all equatorial plane
quantities, while $V_z$ and $F^{-}$ are quantities at the disk's
surface. Additional definitions and relations of these quantities
are
\begin{eqnarray}
   \rho&=&\frac{\Sigma}{H},\\
   l_K&=&\sqrt{\frac{GM_{BH}r^3}{(r-r_g)^2}},\\
   c_s&=&\sqrt{\frac{p}{\rho}},\\
   \Omega&=&\frac{l}{r^2},\\
   F^{-}&=&\frac{24\sigma T^4}{3\tau_R/2+\sqrt{3}+1/\tau_P},\label{eq-bridge}\\
   p&=&k\rho T+p_{rad},\\
   p_{rad}&=&\frac{F^{-}}{12c}\left(\tau_R+\frac{2}{\sqrt{3}}\right),\\
   \tau_R&=&0.34\Sigma(1+6\times10^{24}\rho T^{-3.5}),\\
   \tau_P&=&\frac{1.24\times10^{21}\Sigma\rho T^{-3.5}}{4\sigma},\\
   \beta&=&\frac{k\rho T}{p},
\end{eqnarray}
where equation (\ref{eq-bridge}) is a bridging formula that is
valid for both optically thick and thin regimes, $p_{rad}$ is the
radiation pressure, and $\tau_R$ and $\tau_P$ are the Rosseland
and Planck mean optical depths.

\subsection{Specific Techniques}
As in SM01, in our code the inner edge of the grid is set at
$r\simeq2.5r_g$, close enough to the central black hole so that
the transonic point can be included in the solution; and the outer
boundary is set at $r\simeq 10^4r_g$, far enough away so that no
perturbation from the inner regions could reach. A stationary
transonic disk solution calculated with the $\alpha p$ viscosity
prescription is used as the initial condition for the evolutionary
computations. The $\alpha p$ viscosity prescription may seem
inconsistent with the evolutionary computations adopting the
diffusive viscosity prescription, but this does not matter. In
fact, the initial condition affects only the first cycle of the
obtained limit-cycle solutions, and all following cycles are
nicely regular and repetitive. The time-step $\Delta t$ is
adjusted also in the same way as in SM01 to maintain numerical
stability. We emphasize some techniques specific to our numerical
algorithm below.

The solution to be obtained covers so wide a range of the whole
computational domain, and in particular, the thermal instability
causes a violent variation of the solution (the stiff problem) in
the inner region (inside $200r_g$) of the domain. In such
circumstances the standard one-domain spectral method is certainly
insufficient. Accordingly, we divide the whole computational
domain into $6$ subdomains and let each subdomain contain $65$
grid points, so the total number of grid points is
$65\times6-5=385$ (there are $5$ overlapping point at the
interfaces of subdomains). At each time-level we apply the
one-domain spectral method described in \S\ref{psedo} into each
subdomain. In doing this, various techniques are used to remove or
restrain spurious non-linear oscillations and to treat properly
the boundary conditions, in order to have a numerically stable
solution in each subdomain as well as a stable connection at each
interface. Then we use the scheme described in \S\ref{time-discre}
to carry out the time-integration over the time-step $\Delta t$ to
reach the next time-level.

In general, spurious oscillations are caused by three factors: the
aliasing error, the Gibbs phenomenon, and the absence of viscous
stress tensor components in the basic equations. The aliasing
error is a numerical error specific to the spectral method when it
is used to solve differential equations that contain non-linear
terms, and can be resolved by the spectral filtering technique
described in Appendix (see \citealt{peyret2002} for a detailed
explanation and \citealt{Chan2005} for a quite successful
application of this technique).

However, the spectral filtering itself cannot resolve the Gibbs
phenomenon characteristic to the stiff problem, and the adaptive
domain decomposition method described in \S\ref{domain-decom}
becomes crucially important. In our computations, we set the
spatial location where a large variation appears (e.g., the peak
of the surface density $\Sigma$) as the interface of two certain
subdomains and use the mapping equation (\ref{adjmapp}) along with
the connection conditions (\ref{connect}) and (\ref{connect2}), so
that more grid points are concentrated on the two sides of the
interface to enhance the resolution, and a stable connection
between the two subdomains is realized. As the location of large
variation is not fixed and instead shifts during time-evolution,
we follow this location, redivide the computational domain and
recalculate the mapping parameter for each new subdomain with
equation (\ref{para-a}). In practice, the stiff problem appears
and its location shifts during the first about $30$ seconds of
each cycle whose whole length is $\sim700$ seconds. In these $30$
seconds we have to redivide the domain and reset the grid after
every interval of $0.1$ seconds (or every $0.01$ seconds for the
first a few seconds), and the typical length of time-step $\Delta
t$ is about $10^{-6}-10^{-7}$ seconds. For the rest time of a
cycle (more than $600$ seconds), the stiff problem ceases, then
the grid reset is not needed and the time-step can be much longer.

In addition to the above two factors in the numerical sense, there
is a factor in the physical sense that can also cause spurious
oscillations. The viscous stress tensor has nine spatial
components, but in 1-dimensional studies usually only the $r\phi$
component is included in the basic equations, and omitting other
components can result in numerical instabilities. In particular,
\citet{Szuszkiewicz1997} noted that if the tensor components
$\tau_{rr}$ and $\tau_{\phi\phi}$ were neglected from the radial
momentum equation, some instability would develop and cause
termination of the computation because of the non-diffusive nature
of the equation; and that the instability was suppressed when a
low level of numerical diffusion was added artificially into the
equation. In our code we resolve the similar problem in a slightly
different way. We add into the radial momentum equation
(\ref{eq-radial}) two viscous forces $F_{rr}$ and $F_{\phi\phi}$
in the form as given by equation (5) of \citet{Szuszkiewicz1997},
and accordingly add into the energy equation (\ref{eq-energy}) a
heating term due to viscous friction in the radial direction as
given by equation (15) of \citet{Szuszkiewicz1997}. We find by
numerical experiments that when a very small viscosity in the
radial direction is introduced, i.e., with the radial viscosity
parameter $\alpha_r\simeq0.05\alpha$, where $\alpha$ is the
viscosity parameter in the $\alpha p$ viscosity prescription, the
spurious oscillations due to the absence of viscous stress tensor
components disappear and the solution keeps nicely stable.

Solving partial differential equations in a finite domain usually
requires the Dirichlet boundary condition, i.e., changing the
values of physical quantities at the boundary points to be their
boundary values supplied a priori, or/and the Neumann boundary
condition, i.e., adjusting the values of derivatives of physical
quantities at the boundary points to be their boundary values
supplied a priori. For spectral methods, \citet{Chan2005}
introduced a new treatment, namely the spatial filter as described
in the Appendix, in order to ensure that the Dirichlet or/and the
Neumann boundary condition can be imposed and numerical
instabilities due to boundary conditions are avoided. We note,
however, that their spatial filter treatment is applicable only
for physical quantities whose boundary derivatives are equal, or
approximately equal, to zero (e.g., the radial velocity $v_r$, its
spatial derivative at the outer boundary is nearly zero); while
for quantities whose boundary derivatives are not negligible
(e.g., the specific angular momentum $l$, its spatial derivative
at the outer boundary is not very small), the boundary treatment
in finite difference methods still works, i.e., to let directly
the values of those quantities at the boundary points be the
specified boundary values.

SM01 supplied boundary conditions at both the inner and outer
edges of the grid, i.e., at the inner edge
$Dl/Dt=(\partial/\partial r+v_r\partial/\partial r)l=0$ and
$\partial p/\partial r=0$, and at the outer edge both $v_r$ and
$l$ are constant in time and $l$ is slightly smaller than the
corresponding $l_K$. In our computations we find that it is not
necessary to supply a priori the two inner boundary conditions as
in SM01. With the two outer boundary conditions as in SM01 and
free inner boundary conditions instead, we are able to obtain
numerically stable solutions of the basic equations, and these
solutions automatically lead to an almost zero viscous torque,
i.e., $Dl/Dt\simeq0$, and a nearly vanishing pressure gradient,
i.e., $\partial p/\partial r\simeq0$, in the innermost zone. This
proves that the inner boundary conditions assessed in SM01 are
correct, but we think that our practice is probably more natural
and more physical. Once the state of an accretion flow at the
outer boundary is set, the structure and evolution of the flow
will be controlled by the background gravitational field of the
central black hole, and the flow should adjust itself in order to
be accreted successfully into the black hole. In particular, both
the viscous torque and the pressure gradient of the flow must
vanish on the black hole horizon, and in the innermost zone a
correct solution should asymptotically approach to such a state,
thus in the computations no inner boundary conditions are
repeatedly needed.

\subsection{Numerical Results}
\begin{figure}
\plotone{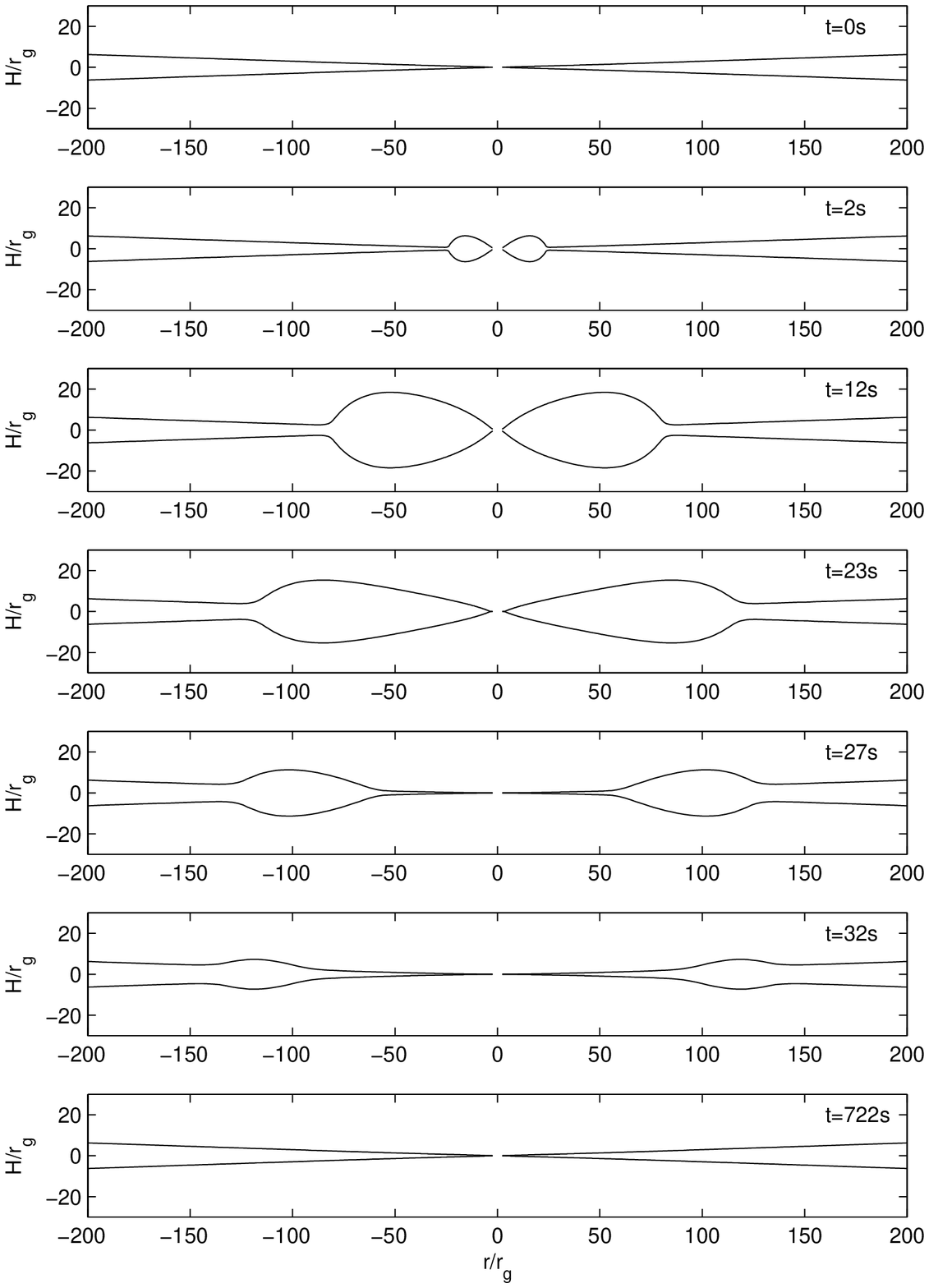} \caption{Evolution of the half-thickness of the
disk during one full cycle.} \label{fig1}
\end{figure}

\begin{figure}
\plotone{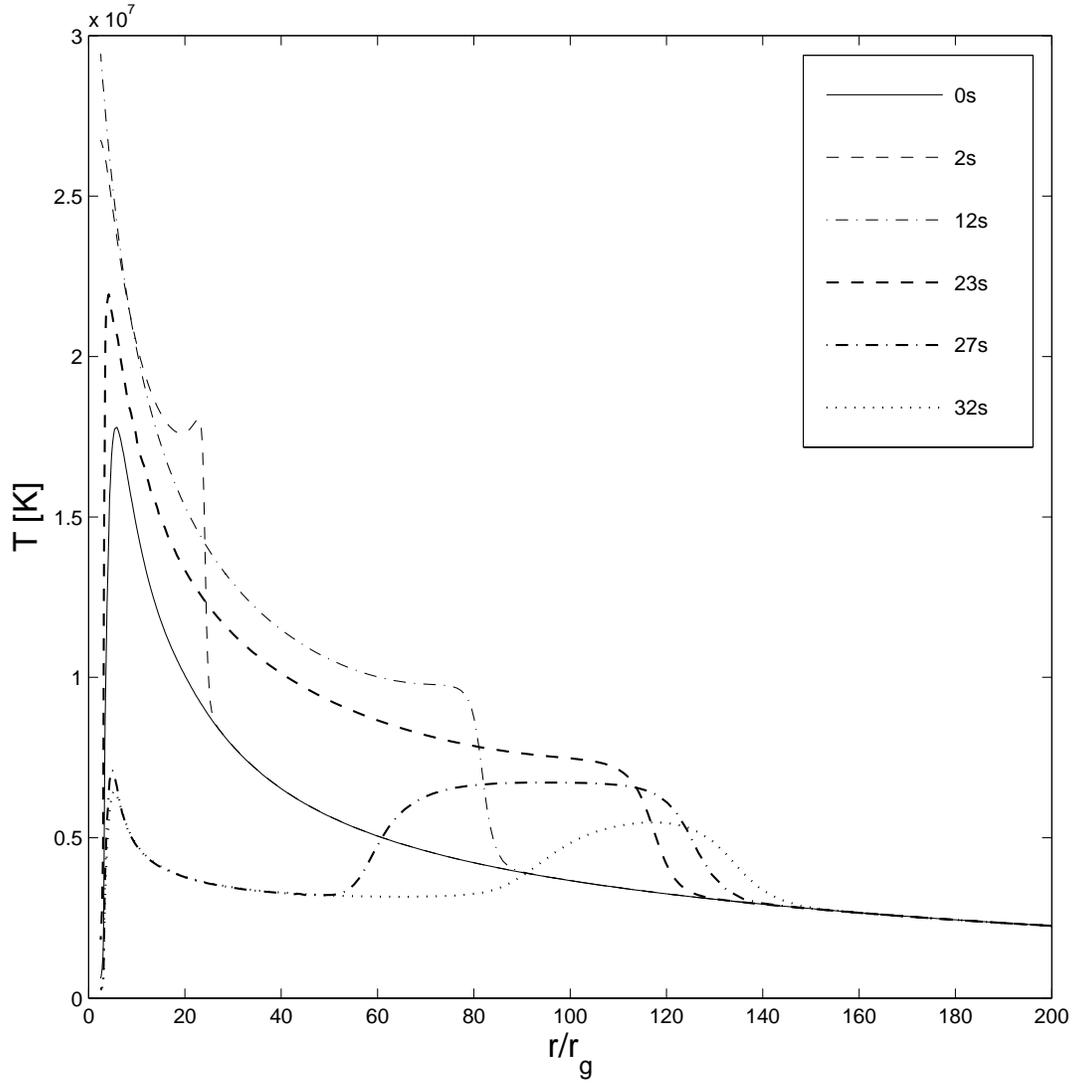} \caption{Evolution of the temperature of the
matter in the disk during one full cycle.} \label{fig2}
\end{figure}

\begin{figure}
\plotone{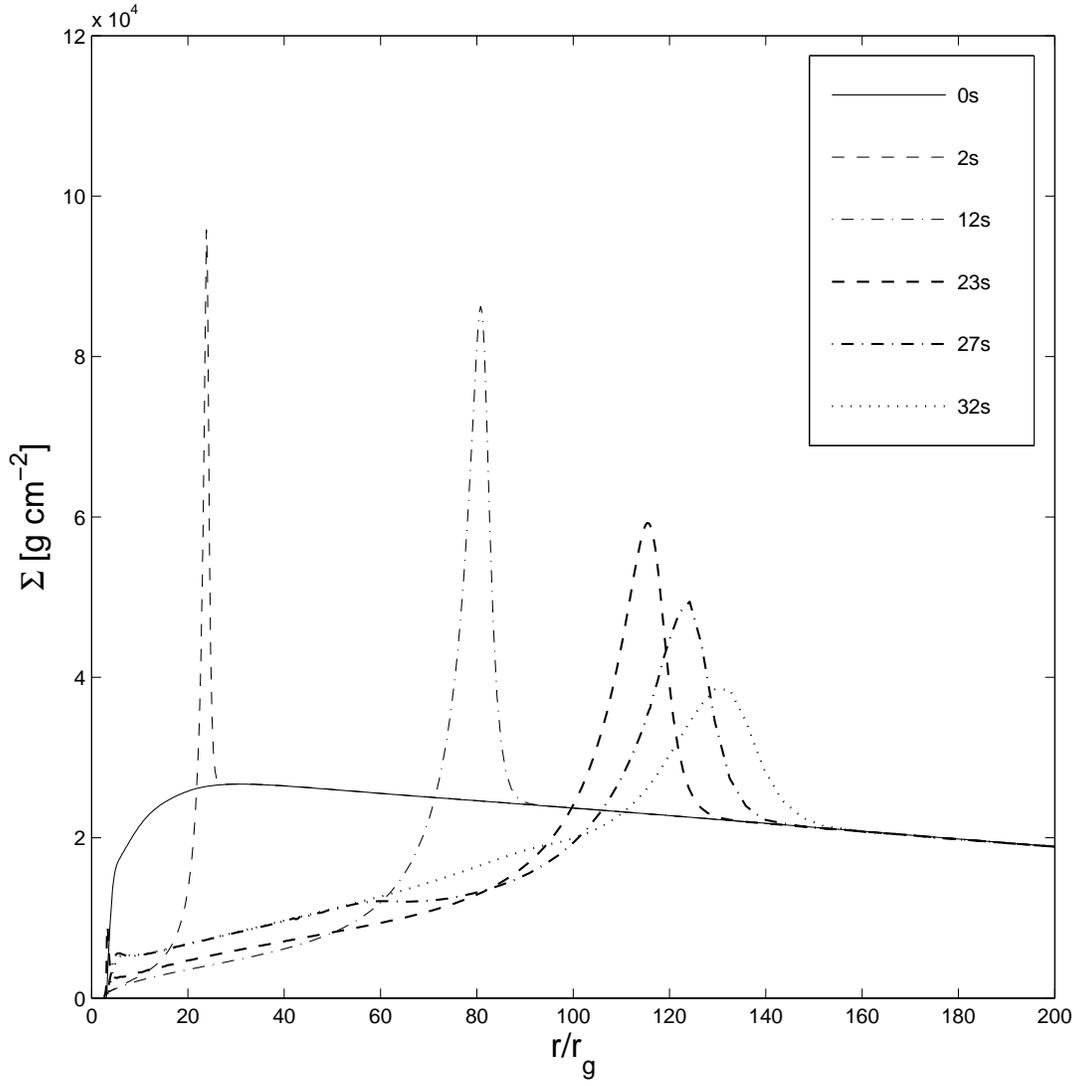} \caption{Evolution of the surface density of the
disk.} \label{fig3}
\end{figure}

\begin{figure}
\plotone{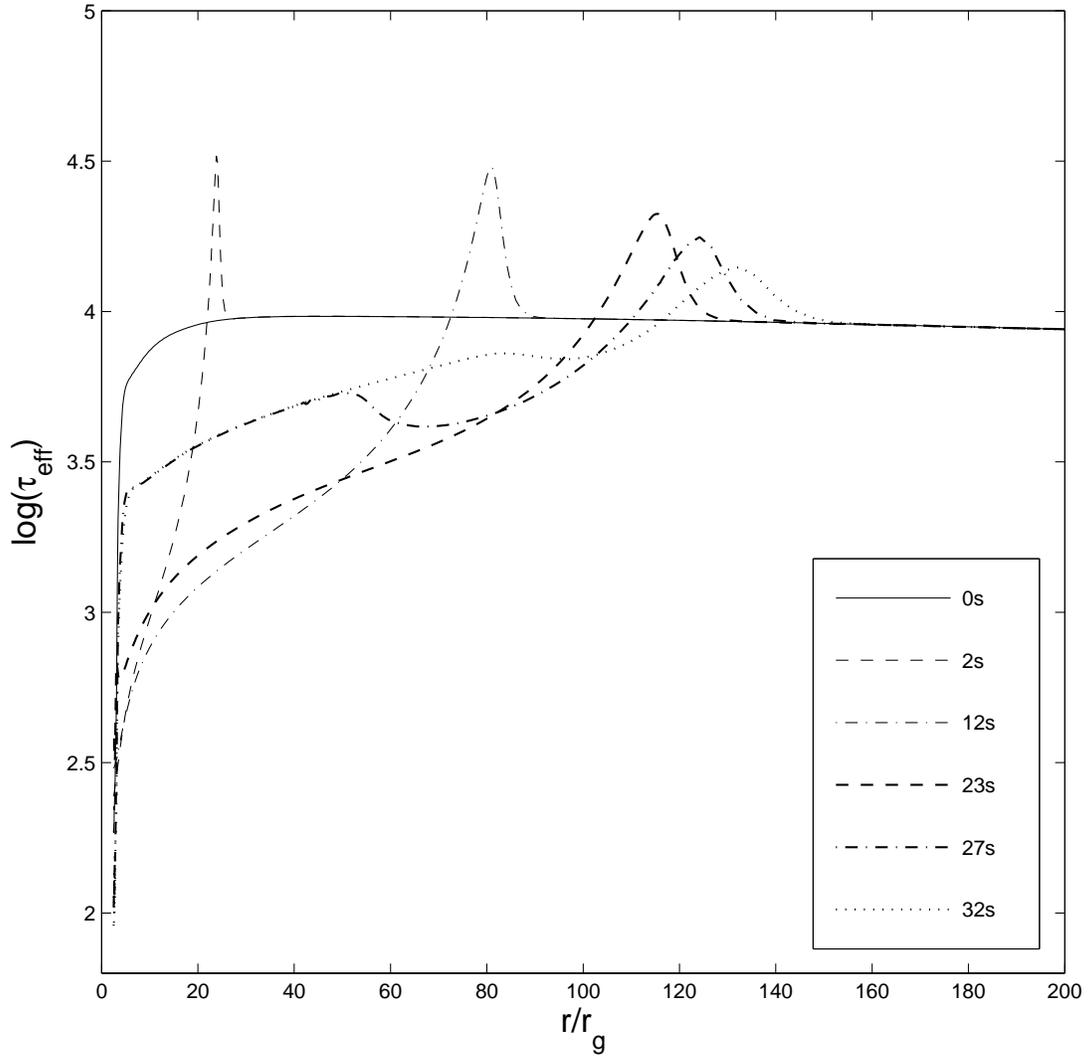} \caption{Evolution of the effective optical depth
of the disk.} \label{fig4}
\end{figure}

\begin{figure}
\plotone{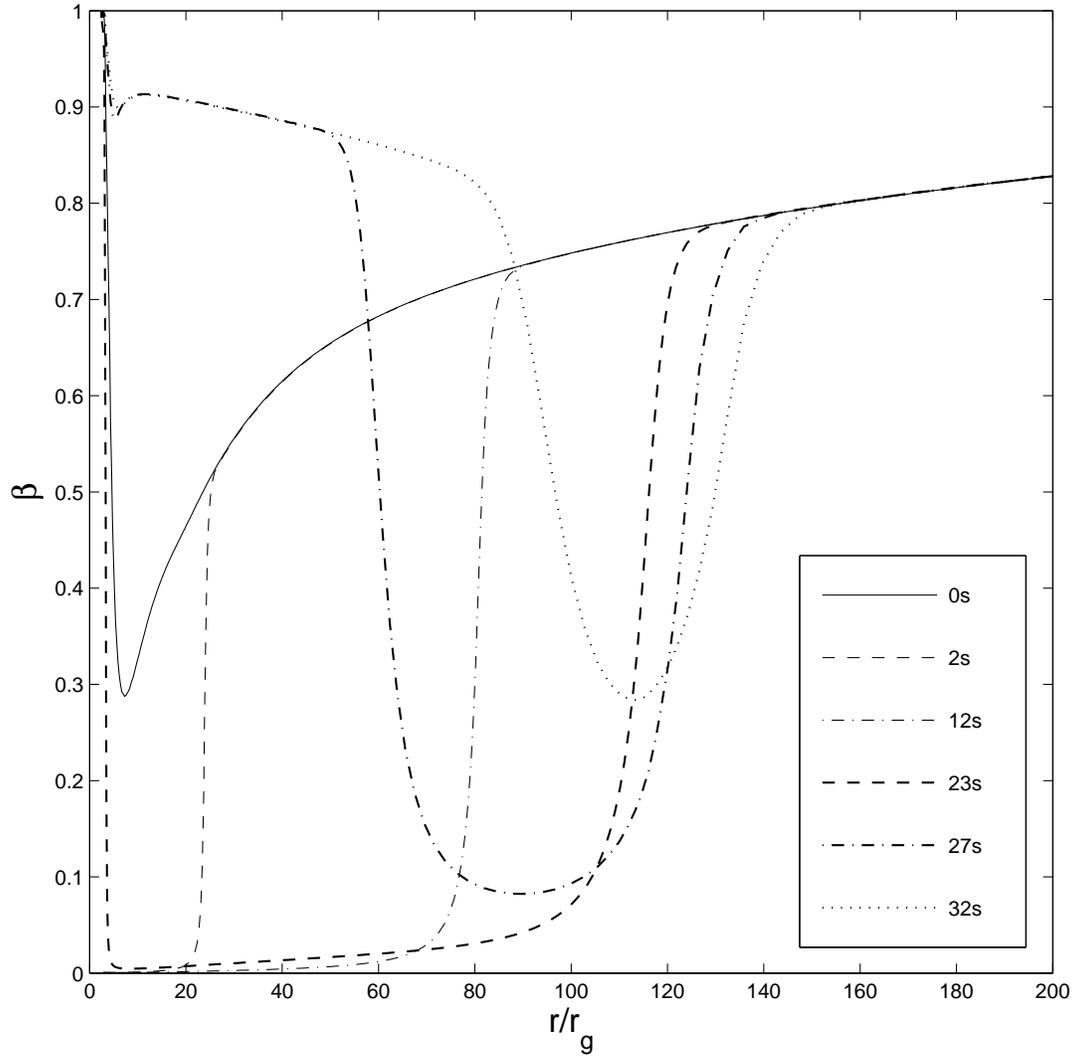} \caption{Evolution of the ratio of gas pressure
to total pressure in the disk.} \label{fig5}
\end{figure}

\begin{figure}
\plotone{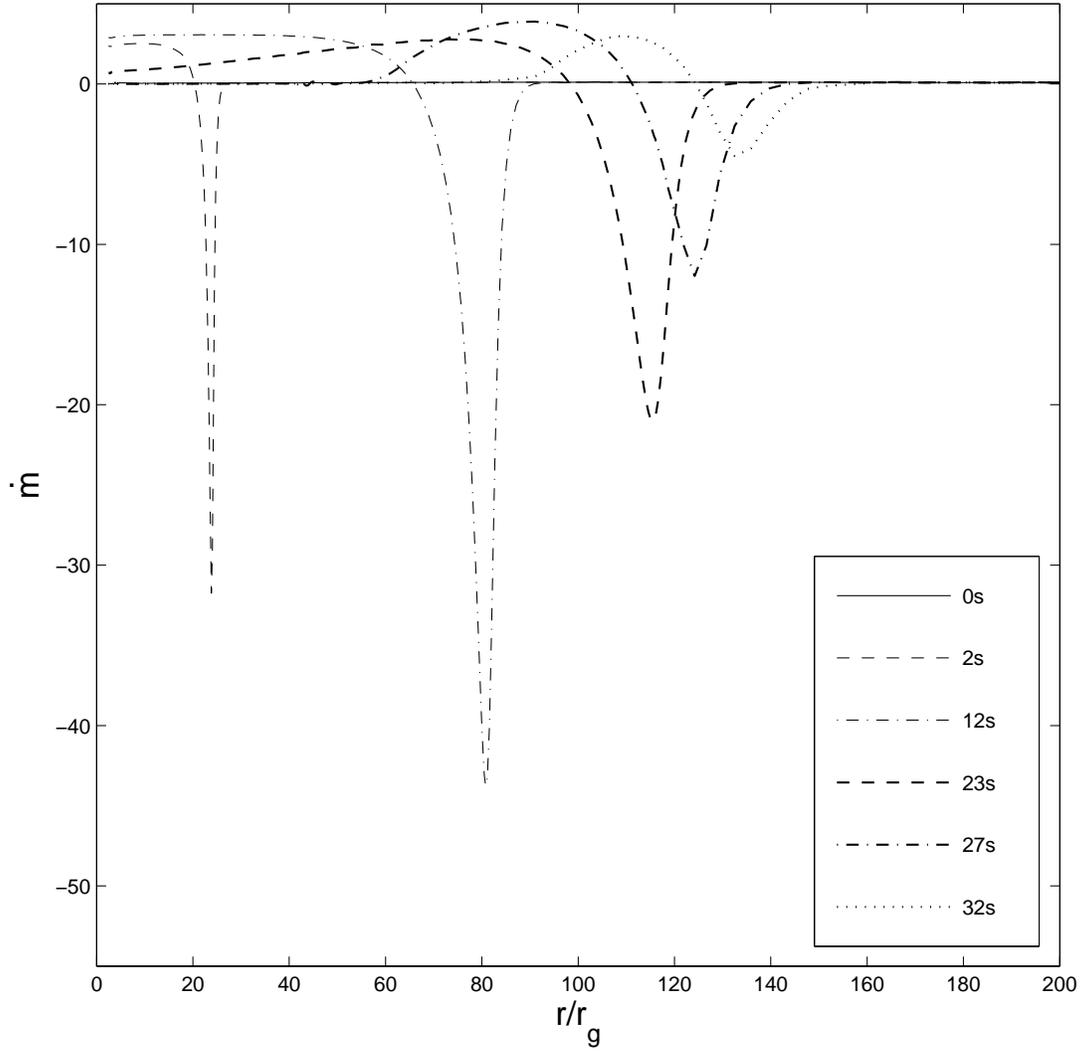} \caption{Evolution of the local accretion rate in
the disk.} \label{fig6}
\end{figure}

We have performed computations for a model accretion disk with
black hole mass $M_{BH}=10M_{\odot}$, initial accretion rate
$\dot{m}\equiv\dot{M}/\dot{M}_{cr}=0.06$ ($\dot{M}=-2\pi r \Sigma
v_r$ is the accretion rate and $\dot{M}_{cr}$ is the critical
accretion rate corresponding to the Eddington luminosity), and
viscosity parameter $\alpha=0.1$ (the equivalent $\alpha$ in the
diffusive viscosity prescription is
$0.1\times(2/3\sqrt{6})\simeq0.0272$). It is known that the inner
region of a stationary accretion disk with such physical
parameters is radiation pressure-supported ($\beta<0.4$) and is
thermally unstable, and the disk is expected to exhibit the
limit-cycle behavior \citep[][Chaps. 4 and 5]{Kato98}. We have
continued computations for several complete limit-cycles, and a
representative cycle is illustrated in Figures \ref{fig1} -
\ref{fig6}, which are for the time evolution of the radial
distribution of the half-thickness of the disk $H$, temperature
$T$, surface density $\Sigma$, effective optical depth
$\tau_{eff}=(2/3)(3\tau_R/2+\sqrt{3}+1/\tau_P)$, ratio of gas
pressure to total pressure $\beta$, and accretion rate $\dot{m}$,
respectively. Note that negative values of $\dot{m}$ signify an
outflow in the radial direction, not in the vertical direction as
the word 'outflow' in the literature usually means.

The first panel of Figure \ref{fig1} and the solid lines in
Figures \ref{fig2} - \ref{fig6} show the disk just before the
start of the cycle ($t=0s$). The disk is essentially in the SSD
state, i.e., it is geometrically thin ($H/r\ll1$) and optically
thick ($\tau_{eff}\gg1$) everywhere, its temperature has a peak at
$r\simeq6r_g$, its accretion rate is nearly constant in space, and
its inner region (from $\sim5r_g$ to $\sim14r_g$) has $\beta<0.4$
and is thermally unstable. Note that this configuration is not a
stationary state and is with the diffusive viscosity prescription,
so it is very different from the initial condition at the
beginning of the computation, which is a stationary solution with
the $\alpha p$ viscosity prescription.

As the instability sets in ($t=2s$, the second panel of Fig.
\ref{fig1} and the thin dashed lines in Figs. \ref{fig2} -
\ref{fig6}), in the unstable region ($r<24r_g$) the temperature
rises rapidly, the disk expands in the vertical direction, a very
sharp spike appears in the surface density profile and accordingly
in the optical depth and accretion rate profiles (exactly the
stiff problem). The spikes move outwards with time, forming an
expansion wave, heating the inner material and pushing it into the
black hole, and perturbing the outer material to departure from
the SSD state. The expansion region is in fact essentially in the
state of slim disk, as it is geometrically thick ($H/r\lesssim1$),
optically thick, very hot, and radiation pressure-supported
($\beta<0.4$); and the front of the expansion wave forms the
transition surface between the SSD state and the slim disk state.
At $t=12s$ (the third panel of Fig. \ref{fig1} and the thin
dot-dashed lines in Figs. \ref{fig2} - \ref{fig6}), in the
expansion region $H$ and $\dot{m}$ (negative, radial outflow)
reach their maximum values, and the local $\dot{m}$ (positive,
inflow) exceeds $3$ which is far above the initial value
$\dot{m}=0.06$ and is even well above the critical value
$\dot{m}=1$.

Once the wavefront has moved beyond the unstable region
($r\lesssim120r_g$), the expansion starts to weaken, the
temperature drops in the innermost part of the disk and the
material there deflates ($t=23s$, the fourth panel of Fig.
\ref{fig1} and the thick dashed lines in Figs. \ref{fig2} -
\ref{fig6}). Subsequently, deflation spreads out through the disk,
and the disk consists of three different parts: the inner part is
geometrically thin, with the temperature and surface density being
lower than their values at $t=0s$; the middle part is what remains
of the slim disk state; and the outer part is still basically in
the SSD state ($t=27s$, the fifth panel of Fig. \ref{fig1} and the
thick dot-dashed lines in Figs. \ref{fig2} - \ref{fig6}).

The 'outburst' part of the cycle ends when it has proceeded on the
thermal time-scale ($t=32s$, the sixth panel of Fig. \ref{fig1}
and the dotted lines of Figs. \ref{fig2} - \ref{fig6}). What
follows is a much slower process (on the viscous time-scale) of
refilling and reheating of the inner part of the disk. Finally
($t=722s$, the seventh panel of Fig. \ref{fig1} and again the
solid lines of Figs. \ref{fig2} - \ref{fig6}), the disk returns to
essentially the same state as that at the beginning of the cycle.
Then the thermal instability occurs again and a new cycle starts.

\begin{figure}
\plotone{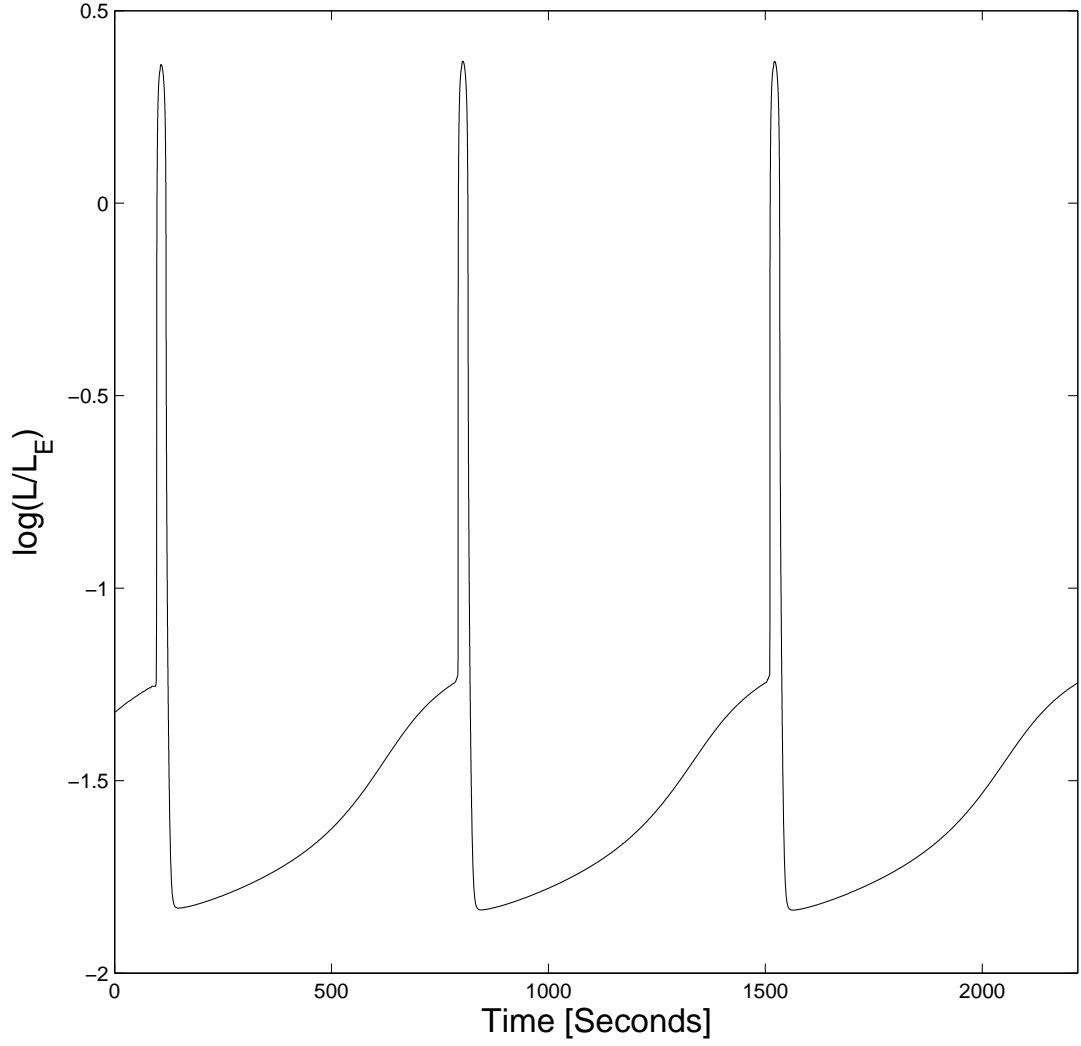} \caption{Variation of the bolometric luminosity
of the disk during three full cycles.} \label{fig7}
\end{figure}

The bolometric luminosity of the disk, obtained by integrating the
radiated flux per unit area $F^{-}$ over the disk at successive
times, is drawn in Figure \ref{fig7} for three complete cycles.
The luminosity exhibits a burst with a duration of about $20$
seconds and a quiescent phase lasting for the remaining about
$700$ seconds of the cycle. The amplitude of the variation is
around two orders of magnitude, during the outburst a
super-Eddington luminosity is realized.

All these results obtained with our numerical method are similar
to that of SM01, not only in the sense that the limit-cycle
behavior of thermally unstable accretion disks is confirmed, but
also in the sense that the numerical solutions are of very good
quality. In our computations we have been able to suppress all
numerical instabilities and to remove all spurious oscillations,
so that in our figures all curves are perfectly continuous and
smooth and all spikes are well-resolved.

\begin{figure}
\plotone{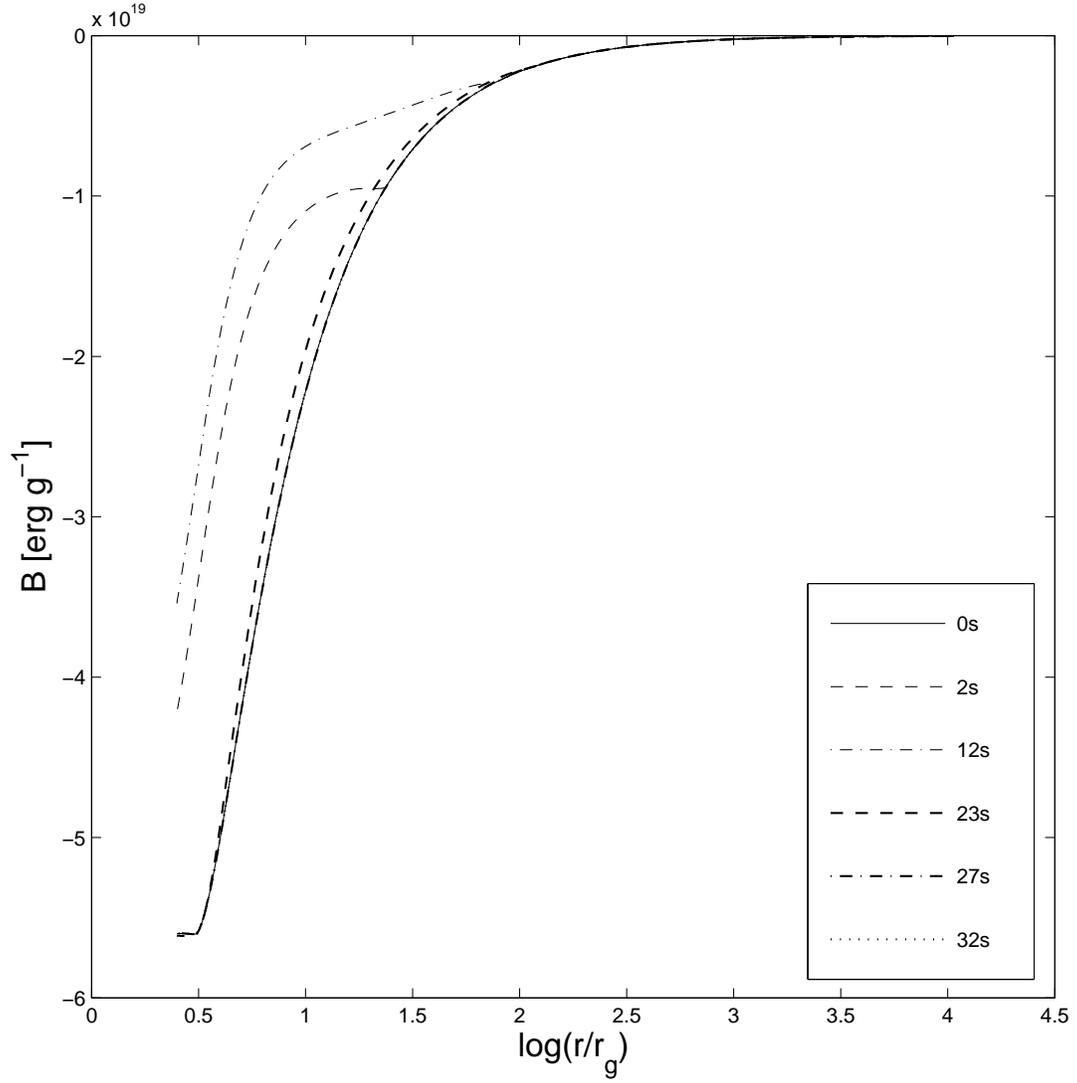} \caption{Bernoulli function of the matter of the
disk. Note that the horizonal scale is very different from that in
Figs. \ref{fig1} - \ref{fig6}.} \label{fig8}
\end{figure}

What is new, however, is that we have also computed the Bernoulli
function (i.e., the specific total energy) of the accreted matter
that is expressed as (cf. Eq. [11.33] of \citealt{Kato98})
\begin{equation}\label{Bernu}
B=\left[3(1-\beta)+\frac{\beta}{\gamma-1}\right]\frac{p}{\rho}+\frac{1}{2}\left(
v_r^2+V_z^2+\Omega^2r^2\right)-\frac{GM_{BH}}{\sqrt{r^2+H^2}-r_g}
\end{equation}
where $\gamma$ is the specific heat ratio and is taken to be
$5/3$. Figure \ref{fig8} shows the quantity $B$ obtained in the
whole computational domain ranging from $r\simeq2.5r_g$ to
$r\simeq10^4r_g$. It is clear that $B$ has negative values for the
whole spatial range (approaching to zero for very large $r$) and
during the whole time of the cycle (in the figure the thick
dot-dashed line for $t=27s$ and the dotted line for $t=32s$ are
coincided with the solid line for $t=0s$). Note that in equation
(\ref{Bernu}) the vertical kinetic energy $0.5V_z^2$ is included,
and the gravitational energy is that for the surface of the disk.
If the vertical kinetic energy is omitted and the gravitational
energy is taken to be its equatorial plane value as in
1-dimensional models, then $B$ will have even larger negative
values. This result is in strong support of the analytical
demonstration of \citet{Abr00} that accretion flows with not very
strong viscosity ($\alpha\lesssim0.1$) have a negative Bernoulli
function; and implies that outflows are unlikely to originate from
thermally unstable accretion disks we consider here, because a
positive $B$ is a necessary, though not a sufficient, condition
for the outflow production.

\section{Summary and Discussion}
We have introduced a numerical method for studies of thermally
unstable accretion disks around black holes, which is essentially
a combination of the standard one-domain pseudo-spectral method
\citep{Chan2005} and the adaptive domain decomposition method
\citep{bayliss1995}. As a test of our method, for the case of
moderate viscosity we have reproduced the best numerical results
obtained previously by SM01. Despite these similarities, we have
made the following improvements over previous works in the
numerical algorithm and concrete techniques, which have been
proven to be effective in the practice of our computations.

1. In applying the domain decomposition method to resolve the
stiff problem, we develop a simple and useful connection technique
to ensure a numerically stable continuity for the derivative of a
physical quantity across the interface of two conjoint subdomains,
i.e., equations (\ref{connect}) and (\ref{connect2}), instead of
the connection technique in \citet{bayliss1995} that is seemingly
complicated and was not explicitly explained.

2. We construct a mapping function (eq. [\ref{adjmapp}]) by adding
the simple linear mapping function (eq. [\ref{trivialmap}]) into
the adaptive mapping function (eq. [\ref{adjmapp2}]) proposed by
\citet{bayliss1995}, so that the mapping between the
Chebyshev-Gauss-Lobatto collocation points $\bar{r}_k$ and the
physical collocation points $r_k^{(i)}$, not only the mapping
between two sets of collocation points $\bar{r}_k$ and
$\tilde{r}_k$, is completed; and the adjustability of equation
(\ref{adjmapp2}) is kept to enable us to follow adaptively the
region that is with the stiff problem and is shifting in space
during time-evolution.

3. For the time-integration, we use two complementary schemes,
namely the third order TVD Runge-Kutta scheme and the third order
backward-differentiation explicit scheme. The former scheme is
popular in one-domain spectral methods and is essentially what was
used by \citet{Chan2005}, and the latter one can achieve the same
accuracy and has advantage of lower CPU-time consumption.

4. For the treatment of boundary conditions, we notice that the
spatial filter technique developed by \citet{Chan2005} for
spectral methods is useful but is itself alone insufficient, and
the treatment traditionally used in finite difference methods is
still needed to complement. We also find that once reasonable
conditions are set at the outer boundary, our solutions behave
themselves physically consistent close to the black hole horizon,
and no inner boundary conditions are necessary as supplied by
SM01.

5. We resolve the problem of spurious oscillations due to the
absence of viscous stress tensor components in the basic equations
in a way different from that of SM01. SM01 introduced an
artificial viscous term in the radial and vertical momentum
equations. We instead improve the basic equations by including two
viscous force terms $F_{rr}$ and $F_{\phi\phi}$ in the radial
momentum equation and a corresponding viscous heating term in the
energy equation, all these terms were already mentioned by the
same authors of SM01 in an earlier paper \citep{Szuszkiewicz1997}.
As for the vertical momentum equation, because of its crudeness in
our $1.5$-dimensional studies, we still adopt an artificial term
whose explicit form is kindly provided by Szuszkiewicz \& Miller
and is unpublished. We obtain solutions at the same quality level
as in SM01, but we think that our treatment is probably more
physical in some sense. In particular, any modification in the
momentum equation ought to require a corresponding modification in
the energy equation, otherwise the energy conservation would not
be correctly described.

Of these five improvements, we expect that the first two and the
last one will be particularly helpful for our subsequent studies
of the strong viscosity case ($\alpha\sim1$). In this case the
viscous heating becomes extremely huge, the 'outburst' of the disk
due to the thermal instability is predicted to be more violent,
and the Gibbs phenomenon related to the stiff problem can be even
more serious than in the case of moderate viscosity studied in
this paper. Our improved domain decomposition method is prepared
to front these difficulties. As for another nettlesome problem
that the absence of some viscous stress tensor components in $1$-
or $1.5$-dimensional equations can also cause serious spurious
oscillations, we think that, although in the moderate viscosity
case they are equivalently effective as what were made by SM01,
our modifications for both the radial momentum and energy
equations will show their advantages in the strong viscosity case.
In fact, the importance of the viscous forces $F_{rr}$ and
$F_{\phi\phi}$ has long since been pointed out
\citep[e.g.,][]{papaloizou86}. We think that the inclusion of a
heating term in the energy equation in accordance with these two
forces will be not only consistent in physics, but also hopefully
important in obtaining numerically stable solutions. With all
these preparations made in this paper, we wish to achieve the goal
to answer the question of the fate of thermally unstable black
hole accretion disks with very large values of $\alpha$: do these
disks finally form stable and persistent SSD+ADAF configurations
as suggested by \citet{TM98}, or they also undergo limit-cycles,
or something else? In view of the two facts that limit-cyclic
luminosity variations, even though with seemingly very reliable
theoretical warranties, are not usually observed for black hole
systems (GRS 1915+105 remains the only one known); and that
outflows are already observed in many high energy astrophysical
systems that are believed to be powered by black hole accretion,
but are unlikely to originate from thermally unstable accretion
disks we study here because of the negative Benoulli function of
the matter in these disks, it will be definitely interesting if
some behaviors other than the limit-cycle for non-stationary black
hole accretion disks and/or the outflow formation from these disks
can be demonstrated theoretically.

\acknowledgments

We are very grateful to Ewa Szuszkiewicz and John C. Miller for
many helpful instructions and providing an unpublished formula of
the numerical viscosity. We also thank Wei-Min Gu for beneficial
discussions. This work was supported by the National Science
Foundation of China under grant 10673009.

\appendix

\section{Spectral Filtering and Boundary Conditions}
When applied to solve non-linear partial differential equations, a
principle drawback of spectral methods is the aliasing error that
can cause  spurious oscillations at high frequencies. The spectral
filtering is a special technique developed to filter out the
high-frequency modes in each time-step to reduce the aliasing
error. As in \citet[][see~also
\citealt{gottlieb1997,peyret2002}]{Chan2005}, we use a exponential
filter in spectral space as
\begin{equation}
    \sigma_{\delta}\left(\frac{n}{N}\right)\equiv\exp\left(-|\ln\epsilon|\left|\frac{n}{N}\right|^{\delta}\right),
\end{equation}
where $\epsilon$ is the machine accuracy and $\delta$ is a
parameter that can be determined from the numerical practice. Then
instead of $u(r_k)$, the filtered collocation values of the
physical quantity $u(r)$ at a collocation point $r_k$ are given by
\begin{equation}
   \bar{u}(r_k)=\sum\limits_{n=0}^{N}\sigma_{\delta}\left(\frac{n}{N}\right)\hat{u}_n
    \cos\left(\frac{nk\pi}{N}\right).
\end{equation}

As for the boundary conditions, the Dirichlet condition,
$u|_{boundary}=u_0$, or/and the Neumann condition, $(\partial
u/\partial r)|_{boundary}=u_0^{'}$, are generally required. In
order to avoid the appearance of discontinuous step-functions at
the boundaries that would cause the Gibbs Phenomenon,
\citet{Chan2005} introduced another filtering technique, namely
the spatial filter. In our numerical algorithm, we either follow
\citet{Chan2005} to impose the boundary conditions by using the
spatial filter, or directly change the value of a certain physical
quantity to be its given value at the boundary point, depending on
whether the boundary derivative of the quantity is or is not very
small. The spatial filter is a monotonically decreasing filter
\begin{equation}
h(r)=\exp\left[-|\ln\epsilon|\left(\frac{r-r_{min}}{r_{max}-r_{min}}\right)^\delta\right]
\end{equation}
for the outer boundary, and a monotonically increasing filter
\begin{equation}
h(r)=\exp\left[-|\ln\epsilon|\left(\frac{r_{max}-r}{r_{max}-r_{min}}\right)^\delta\right]
\end{equation}
for the inner boundary. With this filter, the Dirichlet boundary
condition can be imposed in each time-step as
\begin{equation}
u_k^n \rightarrow (u_k^n-u_0)h(r_k)+u_0,
\end{equation}
and the Neumann boundary condition can be imposed as
\begin{equation}
\left(\frac{\partial u}{\partial r}\right)_k^n \rightarrow
   \left[\left(\frac{\partial u}{\partial
   r}\right)_k^n-u_0^{'}\right]h(r_k)+u_0^{'},
\end{equation}
where the superscript $n$ and subscript $k$ denote the relevant
values at the $n$-th time-level and the collocation point $r_k$,
respectively.

\clearpage

\end{document}